\documentstyle[12pt,aasms4,epsf]{article}

\def\lapprox{\hbox{\lower .8ex\hbox{$\,\buildrel < \over\sim\,$}}}
\def\gapprox{\hbox{\lower .8ex\hbox{$\,\buildrel > \over\sim\,$}}}
\def\cago{$^{12}$C$(\alpha,\gamma)^{16}$O }

\begin{document}

\title{The halo white dwarf population}

\author{Jordi Isern\altaffilmark{1},
        Enrique Garc\'\i a-Berro\altaffilmark{2}, 
        Margarida Hernanz\altaffilmark{1}, 
        Robert Mochkovitch\altaffilmark{3}, and
        Santiago Torres\altaffilmark{4}
	}
\altaffiltext{1} {Institut for Space Studies of Catalonia--CSIC,
	          Edifici Nexus--104, Gran Capit\`a 2--4, 08034 
		  Barcelona, Spain}
\altaffiltext{2} {Departament de F\'{\i}sica Aplicada, Universitat
                  Polit\`ecnica de Catalunya \& Institut for Space 
		  Studies of Catalonia--UPC, Jordi Girona Salgado s/n,
		  M\`odul B-5, Campus Nord, 08034 Barcelona, Spain}
\altaffiltext{3} {Institut d'Astrophysique de Paris, C.N.R.S., 98 bis 
		  Bd. Arago, 75014 Paris, France}
\altaffiltext{4} {Departament de Telecomunicaci\'o i Arquitectura de
                  Computadors, EUP de Matar\'o, Universitat 
		  Polit\`ecnica de Catalunya, Avda. Puig Cadafalch 101,
                  08303 Matar\'o, Spain}

\received{} 
\accepted{}

\begin{abstract}
Halo white dwarfs can provide important information about the 
properties and evolution of the galactic halo. In this paper we 
compute, assuming a standard IMF and updated models of white dwarf
cooling, the expected luminosity function, both in luminosity and 
in visual magnitude, for different star formation rates. We show 
that a deep enough survey (limiting magnitude $\gapprox 20$) could 
provide important information about the halo age and the duration of 
the formation stage. We also show that the number of white dwarfs 
produced using the recently proposed biased IMFs cannot represent 
a large fraction of the halo dark matter if they are constrained 
by the presently observed luminosity function. Furthermore, we show 
that a robust determination of the bright portion of the luminosity 
function can provide strong constraints on the allowable IMF shapes. 
\end{abstract}


\section{Introduction}

One way to understand the structure and evolution of the Galaxy is to 
study the properties of one of its fossil stars: white dwarfs. Their 
luminosity function has been extensively studied since it provides 
important information about the properties of the Galaxy and new deep 
surveys will open the possibility to observe the white dwarf population 
beyond the cutoff reported by Liebert, Dahn \& Monet (1988) and Oswalt 
et al. (1996) as well as to discriminate, on the basis of their 
kinematical properties, those that belong to the halo. If the halo 
was formed sometime before the disk as a burst of short duration 
(Eggen et al. 1962), it would be possible to obtain information about 
the time elapsed between the formation of both structures (Mochkovitch 
et al. 1990). If mergers of protogalactic fragments have played an 
important role in the formation of the galactic halo (Searle \& Zinn 
1978), their signature should be apparent in the white dwarf luminosity 
function.

The observed properties of halo white dwarfs are very scarce. These 
properties can be summarized as follows: a) Liebert, Dahn \& Monet 
(1989) provided a very preliminary luminosity function using six white 
dwarfs, which were identified as halo members because of their high 
tangential velocities. b) Flynn, Gould \& Bahcall (1996) have found 
that the number of stellar objects in the Hubble Deep Field (HDF) with 
$V-I>1.8$ is smaller than 3, while M\'endez et al. (1996) have 
identified 6 objects with $0<V-I<1.2$ that could be white dwarfs 
(although they recommend to reject them as white dwarf candidates 
because of their colors). Both values can be considered as reliable 
upper limits to the number of white dwarfs in the HDF. c) A recent 
analysis of the microlensing events towards the Large Magellanic Cloud 
suggests that they are produced by halo objects with an average mass 
$\sim 0.5 \, M_{\sun}$ (Alcock et al. 1997) and the total contribution 
of such objects to the total halo mass could be as high as 40\%. 
Obviously, white dwarfs are one of the most natural candidates to 
explain such observations. If this interpretation of the microlensing 
data turns out to be correct, the tight constraints imposed by galactic 
properties should demand the use of biased non-standard initial mass 
functions (Adams \& Laughlin 1996; Fields, Mathews \& Schramm 1996; 
Chabrier, S\'egretain \& M\'era 1996).

These IMFs are characterized by a pronounced shortfall below $\sim 
1\, M_{\sun}$ and above $\sim 7-10 \,M_{\sun}$ in order to avoid the 
overproduction of red dwarfs in the first case, and to avoid problems 
with the luminosity of galactic haloes at high redshift (Charlot \& 
Silk 1995) or to overproduce heavy elements by the explosion of massive 
stars in the second case. The problem is that the IMF determines 
the contribution of each kind of star to galactic evolution and, 
therefore, any ad-hoc change introduced to solve a problem, riscs 
to introduce a desadjustement in other apparently well settled fields. 
In the case of the IMFs quoted here, the problem is twofold. First, 
the quantity of astrated mass that is returned on average to the 
interstellar medium through stellar winds and planetary nebula 
ejection is very high, it lies in the range of 40 to 80\%. If white 
dwarfs contribute substantially to the mass budget of the halo, the 
total energy necessary to eject this unwanted mass to the 
intergalactic medium is $\sim 10^{60}$ erg, assuming a typical 
halo radius of 50 kpc and a mass of $10^{12} \, M_{\sun}$. Since 
the number of massive stars has already been suppressed to avoid 
an excess of supernovae, an important part of this material would 
remain locked into the galaxy (Isern et al. 1997a). Furthermore, 
since intermediate mass stars are the main producers of carbon and 
nitrogen, it would be hard to account for the [C,N/O] ratios observed 
in Population II stars if an important part of this matter is
invested in the formation of these stars (Gibson \& Mould 1997). 
Second, even if these problems were solved, the huge increase of 
white dwarfs would increase the number of Type Ia supernovae unless 
new ad-hoc and completely unjustified hypothesis about the properties 
of binary stars are adopted. The resulting overproduction of Fe and 
the excessively high rate of supernova explosions have recently led 
(Canal, Isern \& Ruiz-Lapuente 1997) to the conclusion that, in all 
cases, the contribution of white dwarfs to the halo mass should be 
well below 5--10 \%. At this point, it is worthwhile to point out 
that both the estimated mass of the objects causing the microlensing 
events and the total mass of the population responsible of such events 
are still preliminary (Mao \& Paczynski 1996; Nakamura, Kan-ya \& 
Nishi 1996, Zaritsky \& Lin 1997).

In view of the interest of the luminosity function of halo white 
dwarfs and since there is not any recent study of such stars using 
standard hypothesis, it is worthwhile to construct an updated
series of standard models of halo white dwarf populations for 
comparison purposes.

\section{The luminosity function}

The luminosiy function is defined as the number of white dwarfs per unit
volume and per unit of bolometric magnitude,$M_{\rm bol}$:

\begin{equation}
n(M_{\rm bol},T)= \int^{M_{\rm s}}_{M_{\rm i}}\, \Phi (M)\, \Psi \big
(T-t_{\rm cool}-t_{\rm MS})\,
\tau_{\rm cool} \;dM
\end{equation}

\noindent
where $M$ is the mass of the parent star (for convenience all white 
dwarfs are labelled with the mass of their main sequence progenitors), 
$\tau_{\rm cool}=dt/dM_{\rm bol}$ is the characteristic cooling time,
$M_{\rm s}$ and $M_{\rm i}$ are the maximum and the minimum masses of
the main sequence stars able to produce a white dwarf of magnitude 
$M_{\rm bol}$ at the time $T$, $t_{\rm cool}$ is the time necessary 
to cool down to this magnitude, $t_{\rm MS}$ is the main sequence 
lifetime and $T$ is the age of the population under study (disk, 
halo\ldots). Of course, for evaluating equation (1) a relationship 
between the mass of the white dwarf and the mass of its progenitor 
must be provided. It is also necessary to provide a relationship 
between the mass of the progenitor and its main sequence lifetime. 
We have used those of Wood (1992) instead of those of Iben \& Laughlin
(1989) as we usually did in previous papers in order to compare 
with Adams \& Laughlin (1996). In order to properly compare with 
the observations, it is desirable to bin this function in intervals 
of magnitude $\Delta M_{\rm bol}$, usually of one or half magnitudes, 
in the following way:

\begin{equation}
\langle n(M_{\rm bol},T)\rangle_{\Delta M_{\rm bol}}
        =\frac{1}{\Delta M_{\rm bol}}
       \int_{M_{\rm bol}-0.5\Delta M_{\rm bol}}^{M_{\rm bol}
       +0.5\Delta M_{\rm bol}}
       n(M_{\rm bol},T) \;dM_{\rm bol}
\end{equation}

\subsection{The cooling sequences}

The values taken by $t_{\rm cool}$ and $\tau_{\rm cool}$ depend on 
the adopted evolutionary models of white dwarfs. Since there have been 
some misunderstandings about the cooling process, it is worthwhile to 
summarize here the most relevant points. After integrating over the 
entire star and assuming that the release of nuclear energy is 
negligible, the energy balance can be written as (Isern et al. 
1997b):

\begin{equation}
L+L_{\nu}=- \int^{M_{\rm WD}}_0 C_{\rm v}\frac{dT}{dt}\,dm
	  - \int^{M_{\rm WD}}_0 T\Bigg(\frac{\partial P}{\partial 
	     T}\Bigg)_{V,X_0}\frac{dV}{dt}\,dm \nonumber\\
	  + \Big(l_{\rm s} + e_{\rm g}\Big) \dot{m}_{\rm c}
\end{equation}

\noindent
where $L$ and $L_{\nu}$ are the photon and the neutrino 
luminosities respectively, $\dot{m}_{\rm c}$ is the rate 
at which the crystallization front moves outwards, and the 
rest of the symbols have their usual meaning. Neutrinos are 
dominant for luminosities larger than $10^{-1} \,L_{\sun}$. 
Nevertheless, since the phase dominated by the neutrino cooling 
is very short and the luminosity function of very bright halo 
white dwarfs is still unknown we start our calculations at 
$10^{-1} \, L_{\sun}$.

All the four terms in the right hand side of this equation depend
on the detailed chemical composition of the white dwarf. We have 
adopted the chemical profiles  of Salaris et al. (1997a,b) for C--O 
white dwarfs (white dwarf masses in the range of 0.5--1 $M_{\sun}$ 
and progenitors in the mass range 0.7--8 $M_{\sun}$), which take into 
account the presence of high quantities of oxygen in the central 
regions due to the high rates of the \cago reaction, and from 
Garc\'{\i}a-Berro et al. (1997) for O--Ne white dwarfs (white 
dwarf masses $\gapprox 1 \,M_{\sun}$ and progenitors in the mass 
range 8--11 $M_{\sun}$).

The first term of the right hand side of equation (3) represents the 
well known contribution of the heat capacity of the star to the total 
luminosity (Mestel 1952). It strongly decreases when the bulk 
of the star enters into the Debye regime (Lamb \& Van Horn 1975). 
The second term takes into account the net contribution of  
compression to the luminosity. It is in general small since the major
part of the compressional work is invested into increasing the Fermi
energy of electrons (Lamb \& Van Horn 1975, Shaviv \& Kovetz 1976). 
The largest contribution to this term comes from the outer, partially 
degenerate layers. In fact, when the white dwarf enters into the
Debye regime, this term can provide in some cases about 80\% of the 
total luminosity preventing in this way the sudden disappearence of the 
star (D'Antona and Mazzitelli 1989). Of course, the relevance and the 
details of this contribution strongly depend on the characteristics 
of the envelope. It is  clear, therefore, that extrapolating the 
behavior of the coldest white dwarfs just assuming that their 
unique source of energy is the heat capacity may not necessarily 
be the best procedure.

The third term in the right hand side represents the energy release 
associated to solidification. The term $l_{\rm s}$ corresponds to
release of the latent heat ($\sim kT_{\rm s}/{\rm nucleus}$, where 
$T_{\rm s}$ is the solidification temperature), and the term 
$e_{\rm g}$ corresponds to the gravitational energy release associated 
to the chemical differentiation induced by the freezing process 
(Mochkovitch 1983, Isern et al. 1997b). It is important to notice 
here that the C--O models of Salaris et al. (1997a,b) predict very 
high oxygen abundances in the central region. This fact minimizes 
the effect of the chemical differentiation as compared with the 
models assuming half carbon and half oxygen homogeneously distributed 
along the star. It is also interesting to notice that this effect 
is almost completely negligible in the case of O--Ne white dwarfs 
(Garc\'{\i}a-Berro et al. 1997).

The main difference between C--O and O--Ne models is that the last 
ones cool down more quickly. For instance, two white dwarfs of 1 
$M_{\sun}$, one made of C--O and the other one of O--Ne take 11.3 
Gyr and 8.1 Gyr respectively to reach a luminosity of $10^{-5}\, 
L_{\sun}$. This is due to the smaller heat capacity and to the 
negligible influence of chemical differentiation settling in O--Ne 
mixtures --- see Garc\'{\i}a-Berro et al. (1997) for a detailed 
discussion.

We would like to stress the importance of the outer layers of the 
white dwarf in the cooling process. They not only can provide  
a major contribution to the total luminosity during the late stages,
as previously stated, but also control the power radiated by the star. 
The difficulties come from the fact that matter is far from ideal 
conditions and that the convective envelope reaches the partially 
degenerate layers. Although our models take into account reasonably 
well the energy released by the compression of the outer layers, the 
rate at which the energy is radiated away remains quite uncertain.

\subsection{Computational procedure}

The white dwarf luminosity function averaged over bins of width
$\Delta M_{\rm bol}$ as given in equation (2) can also be
directly computed in the following way. Assume a population that was
formed according an arbitrary star formation rate $\Psi (t)$. After a
time $T$ since the origin of the Galaxy, the number of white dwarfs
that have a bolometric magnitude in the interval
$M_{\rm bol} \pm 0.5\Delta M_{\rm bol}$ is:

\begin{equation}
N(M_{\rm bol},T)= \int_t \int_M \Phi(M)\,\Psi(t)\,dM\,dt
\end{equation}

\noindent
where the integral is constrained to the domain that satisfies
the condition:

\begin{equation}
     T -t_{\rm cool}(M,M_{\rm bol}-0.5\Delta M_{\rm bol})
 \le t+t_{\rm MS}(M)
 \le T -t_{\rm cool}(M,M_{\rm bol}+0.5\Delta M_{\rm bol})
\end{equation}

\noindent
Dividing this result by $\Delta M_{\rm bol}$ we obtain equation (2).

This expression can be easily computed using standard methods and,
since it does not make use of the characteristic cooling time 
(which demands the use of numerical derivatives to evaluate it) 
it easily allows to obtain the luminosity function in visual 
magnitudes or in any other photometric band.

Figure 1 displays the luminosity function  of a burst of constant star 
formation rate of arbitrary strength and a duration of 0.1 Gyr that 
started 13 Gyr ago obtained using both methods. The upper solid line 
has been obtained using equation (1), the lower lines have been 
obtained by applying the binning procedure of equation (2) to 
equation (1), solid line, and equations (4) and (5), dotted line. 
The last two lines have arbitrarily been shifted downwards by a fixed 
amount to allow comparison. The differences can be considered as 
negligible. It is worthwhile to note here how the sudden rise 
produced by crystallization at $\log(L/L_{\sun}) \simeq -3.7$ 
in the luminosity function (upper curve) is smeared out when 
bins are taken. Note as well that the observational luminosity 
functions are actually derived using such binning. In all cases 
calculations were stoped at $\log (L/L_{\sun})=-5$ to save 
computing time.

\section{Results and discussion}

\subsection{Standard initial mass function}

Figure 2a displays the luminosity functions of halo and disk white 
dwarfs computed with a standard initial mass function (Salpeter 1961). 
The observational data for both the disk and the halo have been taken 
from Liebert et al. (1988, 1989). The theoretical luminosity 
functions have been normalized to the points $\log (L/L_{\sun})\simeq 
-3.5$ and $\log (L/L_{\sun}) \simeq -2.9$ for the halo and the disk 
respectively due to their smaller error bars. The luminosity function 
of the disk was obtained assuming an age of the disk of 9.3 Gyr and a 
constant star formation rate per unit volume for the disk, and those 
of the halo assuming a burst that lasted 0.1 Gyr and started at $t_{\rm 
halo}$= 10, 12, 14, 16 and 18 Gyr respectively. Due to their higher 
cooling rate, O--Ne white dwarfs produce a long tail in the disk 
luminosity function and a bump (only shown in the cases $t_{\rm 
halo}$= 10 and 12 Gyr) in the halo luminosity function. It is 
important to realize here that the faintest white dwarf known, 
ESO 439--26, which has a mass $M=1.1-1.2$ $M_{\sun}$ (Ruiz et al. 
1995) and a luminosity $\log (L/L_{\sun}) \simeq -5$ is clearly an 
O--Ne white dwarf and cannot be a halo white dwarf unless
the halo stopped its star formation activity less than 8 Gyr 
ago, since the time necessary for O--Ne white dwarfs to reach 
this luminosity is at maximum 8 Gyr. In order to make easier the 
comparison with observations, we display in Figure 2b the same 
luminosity function in visual magnitudes. The photometric 
corrections were obtained from the atmospheric tables of
Bergeron et al. (1995). Beyond $M_{\rm V} \gapprox 17$ these 
corrections were obtained by extrapolating those tables. It is 
interesting to notice that the distance between the peaks of the 
halo luminosity functions has increased due to the fact that more 
and more energy is radiated in the infrared as white dwarfs cool 
down. Therefore, the detection of such peaks should allow the
determination of the age of the galactic halo. It is also convenient 
to remark here that the disk white dwarf luminosity function of 
figure 2b was obtained with the age and normalization factor used 
for figure 2a.

If the halo formed from the merging of protogalactic fragments, the 
time scale for halo formation should be larger than 0.1 Gyr and 
therefore, the white dwarf luminosity function should be different 
from those of Figure 2. To show that we have computed the luminosity 
function for bursts that, starting at 12 Gyr, lasted 0.1, 1 and 3 Gyr. 
The last one was inspired by the age distribution of the globular 
cluster sample of Salaris \& Weiss (1997). We see from Figure 3 
that because of the relative lack of sensitivity to the age and 
shape of the star formation rate of the hot portion of the luminosity 
function, the different curves merge when we normalize them to a fixed
observational bin. As a consequence, it is necessary to have precise 
information about the white dwarf population in the region $M_V 
\gapprox 16$ before being able to reach any conclusion.

The shape of the Hertzprung--Russell diagram of halo white dwarfs also
provides useful information about the halo population. Figure 4 
displays the color--magnitude diagram for each one  of the bursts 
of figure 2 using a simulated Monte Carlo sample of 2,000 stars. 
They can be interpreted as the isochrones, including the lifetime 
in the main sequence, of this halo population. The diagram displays 
a characteristic Z--shape produced by the combination of the different 
cooling times of white dwarfs and main sequence lifetimes of their 
progenitors. This feature moves downwards with the age and ultimately 
disappears. Therefore, its detection could provide an indication of 
the halo age.

The duration of the process of formation of the halo is also reflected 
in the color--magnitude diagram. If the halo took a relatively large 
time to form, the Z--feature would cover a large region of the 
color--magnitude diagram and its width could be used as a duration 
indicator if enough white dwarfs with good photometric data were 
available. Figure 5 displays the color--magnitude diagram for a 
burst of constant star formation rate that was 12 Gyr old and 
lasted 3 Gyr.

Another useful quantity is the discovery function. This function gives 
the number of white dwarfs per interval of magnitude which can be 
detected in the whole sky by a survey limited to a given apparent 
magnitude $m_\lambda$ in a photometric band centered at $\lambda$ 
(Mochkovitch et al. 1990). If we limit ourselves to nearby halo 
white dwarfs, this volume can be considered spherical and the 
discovery function, $\Delta_{\rm H}(M_\lambda)$, is readily 
obtained from the luminosity function

\begin{equation}
\Delta_{\rm H}(M_\lambda)= \frac{4 \pi}{3} \,d^3(M_\lambda) 
n_{\rm H}(M_\lambda)
\end{equation}

\noindent
where $d(M_\lambda)$ is the distance at which a white dwarf of absolute 
magnitude $M_\lambda$ has an apparent magnitude $m_\lambda$: 

\begin{equation}
d(M_\lambda)=10^{1+0.2(m_\lambda -M_\lambda)}
\end{equation}

\noindent 
where $d$ is in parsecs. Since white dwarfs with luminosities $\log 
(L/L_{\sun}) \sim -5$ have effective temperatures of $\sim 3000$ K and 
radiate most of their energy in the red or infrared, we have computed 
the discovery function for both the $V$ and $I$ bands assuming in both 
cases a limiting magnitude $M_{\rm V,I} \simeq 20$. For reasonable ages 
of the halo ($t_{\rm halo} \sim$ 12--16 Gyr), the discovery function in 
both the $V$ and $I$ band yield $\sim 500$ stars/magnitude for bright
objects and have a steady decrease with the magnitude. This decrease is
more pronounced for the $V$ band (Figure 6). In any case, the total 
number of stars that we would expect to find in a survey of such 
characteristics is about 1,500 stars. This implies that the average 
number of white dwarfs that we expect to find in a typical Schmidt 
plate of $6^\circ \times 6^\circ$ is about 1.5. One third of them 
should be brighter than magnitude 12.

At this point it is interesting to examine, just as an exercise, the
impact of the recently discovered white dwarf WD~0346+246 (Hambly, 
Smartt \& Hodgkin 1997). The first analysis indicates that this white 
dwarf is placed at a distance $d \sim 40$ pc and it has a tangential 
velocity $v_{\rm T} \sim 250$ km/s, which indicates that it probably 
belongs to the halo. Its absolute visual magnitude is estimated to be 
in the range $ 16.2 \lapprox M_{\rm V} \lapprox 16.8 $. Therefore, if 
we assume that it is the only star of these characteristics within this 
distance, the luminosity function would take the value of $\sim 1.5 
\times 10^{-5}$ mag$^{-1}$pc$^{-3}$ at $M_{\rm V} \approx 16.5$, which 
is in perfect agreement with the results plotted in Figure 2b. Note 
also that the standard halo models could accomodate a density larger 
by a factor $\sim 3$ to that quoted here just assuming that the 
luminosity function has a peak in this region. In this case, the 
halo should be as young as $\sim 11$ Gyr. If the density finally 
turned out to be larger we could start to think about non conventional 
hypothesis. It is also interesting to notice that the expected number 
of white dwarfs per plate with $16 \le M_{\rm I} \le 17$ is in the 
range of $10^{-1}$ to $3 \times 10^{-2}$.

We have also computed the total number of white dwarfs per stereoradian 
in the direction of the Hubble Deep Field assuming a spheroidal
distribution of stars of the kind $\rho (r)=\rho_0 a^2 /(a^2 + r^2)$, 
with $a=2.5$ kpc, a distance of the Sun to the galactic center of 8.5 
kpc, and a limiting apparent visual magnitude $V=26.3$. The total number
of halo white dwarfs in this photometric band goes from a minimum of 
315,000 stars/str for the 10 Gyr burst to 321,000 stars/str for the 
16 Gyr one, while the number of white dwarfs redder than $V-I=1.8$ 
ranges from a minimum of 233 stars/str for the first case to 2,000 
stars/sr for the second case. The number of stars in the window 
$0<V-I<1.2$ takes values in the range 195,000 to 198,000 stars/str. 
This behavior can be easily understood if we note that, due to the 
normalization procedure, the  bright portion of the luminosity 
function is nearly coincident in all cases and that counts limited 
to a given apparent magnitude are dominated by the brightest stars. 
On the contrary, if we limit ourselves to the very red ones, which 
are also the dimmest ones, we are eliminating the brightest white 
dwarfs and, therefore, the number of them in the pencil becomes 
sensitive to the age. Unfortunately, the HDF pencil is so narrow 
($\Delta \Omega = 4.4$ arc min$^2 =3.723\times 10^{-7} $ str) that 
it is impossible to extract from it any valuable information of the 
properties of the different bursts. 

The local density of halo white dwarfs obtained from our luminosity 
functions ranges, depending on the adopted age of the halo, from 
$5.8\times 10^{-5}$ to $1.1\times 10^{-4}$ white dwarfs per pc$^3$ 
(that is from 0.24 to 0.46 white dwarfs in a sphere of 10 pc of radius 
around the Sun). If we assume that the characteristic mass of halo 
white dwarfs is $\sim 0.6$ $M_{\sun}$, this density represents at 
most the 0.6\% of the local dark halo density, 0.01 $M_{\sun}/{\rm 
pc}^3$ (Gilmore 1997). Finally, we have to mention that if we chose 
to normalize to the total density of discovered white dwarfs as in 
Mochkovitch et al. (1990), all the curves will move downwards by 
approximately a factor 0.3 dex.

\subsection{Biased initial mass functions}

In order to account for the MACHO results in terms of an halo white 
dwarf population, Adams \& Laughlin (1996), and Chabrier et al.
(1996) introduced ad-hoc non-standard initial mass functions that 
fall very quickly below $\sim 1$ $M_{\sun}$ and above $\sim 7$ 
$M_{\sun}$. These functions avoid the overproduction of red dwarfs, 
the overproduction of heavy elements by the explosion of massive 
stars (Ryu, Olive \& Silk 1990) and the luminosity excess of the 
haloes of galaxies at large redshift (Charlot \& Silk 1995) and, 
since the formation of very massive and very small stars has been 
inhibited, the proposed IMFs allow to increase the number of white 
dwarfs per unit of astrated mass.

Table 1 displays the mass density in the form of white dwarfs for bursts
of star formation that started at different ages and lasted 0.1 Gyr 
using the IMFs proposed by Adams \& Laughlin (1996), with $m_{\rm c}
=2.3 $ and $\sigma =0.44$ --- AL case --- and the IMFs proposed by 
Chabrier et al. (1996) --- CSM1 and CSM2 cases. The main differences 
between our calculations and those of Adams \& Laughlin (1996) are 
that we have computed the luminosity function without neglecting the 
time spent in the main sequence, we take into account the full effects 
of crystallization and we normalize to the best known bin of the 
observed halo luminosity function instead of trying to reproduce 
a given density of white dwarfs in the halo. The main differences 
with the Chabrier et al. (1996) calculations rely on the normalization 
procedure and on the fact that we use realistic carbon--oxygen profiles 
instead of assuming that C--O white dwarfs are made of an homogeneous 
mixture of half carbon and half oxygen. Besides that, we use average 
binned functions.

Concerning the AL case, the maximum densities that can be reached are
smaller than the 10\% of the dark halo for any reasonable age of the
Galaxy. The same happens with the CSM1 case. Only in the CSM2 case white
dwarfs can represent a noticeable fraction of the halo dark matter. In
fact, in the case of an age $\sim 16$ Gyr the halo would be saturated 
with white dwarfs. The differences with Chabrier et al (1996) are 
essentially due to the different normalization procedure used here. 
Therefore, it is clear that a robust determination of the bright 
portion of the halo white dwarf luminosity function would introduce 
strong constraints on the allowed shapes of the IMFs.

Figure 7 displays the luminosity functions obtained with the 
aforementioned IMFs for bursts that started 12 (left panel) and 14
Gyr (right panel) ago (dotted lines), which we believe are realistic 
values for the age of the halo, normalized to the brightest and more 
reliable observational bin. The luminosity function corresponding to 
the standard case is also displayed for comparison (thick solid line). 
As expected, the position of the peak does not change but its height 
increases since the number of main sequence stars below $\sim 1$ 
$M_{\sun}$ is severely depleted. This behavior is due to two different 
effects: i) The non-standard IMFs have been built to efficiently 
produce white dwarfs (0.18, 0.39, 0.53 and 0.44 white dwarfs per 
unit of astrated mass for the standard, AL, CSM1 and CSM2 cases 
respectively and for a burst 14 Gyr old for instance). ii) The time 
that a white dwarf needs to cool down to the luminosity of the 
normalization bin, $\log(L/L_{\sun})=-3.5$, is $\sim 1.8$ Gyr 
and only main sequence stars with masses smaller than 1 $M_{\sun}$ 
are able to produce a white dwarf with such a high luminosity if the 
halo is taken to be older than 12 Gyr. Since the new IMFs have been 
taylored to reduce the number of stars below $\sim 1$ $M_{\sun}$, it 
is necessary to shift the luminosity function to very high values to 
fit the normalization criterion. For instance, the values that the 
different IMFs take at $M=0.98$ $M_{\sun}$, the mass of the main 
sequence star that produces a white dwarf with the aforementioned 
luminosity, are $\Phi_{\rm S}=0.23$, $\Phi_{\rm AL}=0.06$, 
$\Phi_{\rm CSM1}=0.2$, and $\Phi_{\rm CSM2}=0.01$.

Figure 7 also shows that all the luminosity functions, except the one 
obtained from the standard IMF, are well above the detection limit 
of Liebert et al. (1988) (shown as triangles). This is due to the 
normalization condition adopted here. If we had normalized the 
luminosity function to obtain a density of $1.35 \times 10^{-5}$ 
white dwarfs per cubic parsec brighter than $\log(L/L_{\sun})=-4.35$ 
as in Mochkovitch et al. (1990), all the luminosity functions (long 
dashed in Figure 7) would have been shifted downwards and only those
corresponding to the CSM2 case would have remained above the detection 
limit. Note, however, that except for unrealistic ages of the galactic 
halo, these IMFs are not only  unable to provide an important 
contribution to the halo (see Table 1 below), but also to fit
the observed bright portion of the luminosity function of halo white
dwarfs. For instance, no one of the CSM2 cases appearing in Figure 2
of Chabrier et al. (1996) fits the brightest bin. Therefore a robust 
determination of the bright portion of the luminosity function could 
introduce severe constraints to the different allowable IMFs.

As we have already mentioned in section 2, one of the major 
uncertainties is the transparency of the envelopes when white 
dwarfs are very cool. If they turn out to be more opaque than 
the models used here, the cooling would be slowed down and the 
height of the peaks would be consequently increased. On the contrary, 
if the atmospheres turn out to be more transparent, white dwarfs 
could be able to reach, for a given age, smaller luminosities 
and the peaks of Figure 7 would therefore be reduced. There is however 
one limitation: the properties of the envelopes for white dwarfs 
brighter than $\log (L/L_{\sun})\approx -4$ are reasonably well 
known. Consequently, we have checked this behavior by arbitrarily 
increasing the transparency below $\log (L/L_{\sun})\approx -4$.
Adopting the appropriate factor, it is possible to reduce the height
of the peaks of figure 7 below the detection limits. Nevertheless, 
since the properties of the luminosity function at the normalization 
point have not changed, the contributions of white dwarfs to the halo 
mass budget remain the same as those quoted in Table 1. We have also
checked if a change in the initial--final mass relationship (in the 
sense of favouring the formation of massive white dwarfs) could 
reduce the number of bright white dwarfs, but we have only obtained 
a slight change in the morphology of the peak since the bright part 
of the luminosity function is dominated by long lived main sequence
stars.

\section{Conclusions}

We have computed the luminosity function of halo white dwarfs for 
different photometric bands assuming a standard IMF and several star 
formation rates. We have shown that a detailed knowledge of this 
function can provide critical information about the halo properties, 
in particular its age and duration of the process of formation. The 
discovery functions computed in this way show that the luminosity 
function can only be obtained if deep enough, $M \gapprox 20$, surveys 
in the $I$ or $R$ bands are performed.

We have also examined the constraints introduced by the Huble Deep Field
and we have found that it is too narrow to be useful in this issue. The
differences between our results and those of Kawaler (1996) are 
probably due to the fact that we do not neglect the lifetime in 
the main sequence since this assumption is not true for bright 
dwarfs, which are dominant in the star counts below a given 
magnitude.

Finally, we have shown that, even using biased IMFs, it is impossible to
appreciably fill the dark halo with white dwarfs if the luminosity
functions are normalized to the observational bin with the smallest 
error bar. Besides the lack of any physical reason able to justify 
the radical changes introduced in biased IMFs and the secondary effects 
mentioned in the Introduction, it is necessary to assume that white 
dwarfs become very transparent when they cool down below $\log(L/L_
{\sun})\approx -4$ (or that for some reason halo white dwarfs cool 
down more quickly than disk white dwarfs, or that they suffered some
kind of selection effect) in order to have escaped detection during 
previous surveys.

{\em Acknowledgements}
	One of us (J.I.) is very indepted to A. Burkert and J. Truran
	for very useful discussions. This work has been supported by 
	DGICYT grants PB94-0111, PB94--0827-C02-02, by the CIRIT grant 
        GRC94-8001, by the AIHF 1996--106 and by the C$^4$ consortium.

\newpage

\begin{table}
\begin{center}
\caption{Local density of halo white dwarfs ($M_{\sun}$/pc$^3$) for 
different IMFs and ages of the halo}
\label{tbl-1}
\begin{tabular}{cccc} \\
\tableline
\tableline  \\
IMF & 12 Gyr&  14 Gyr & 16 Gyr \\ 
\tableline  \\
Standard&$4.3\times 10^{-5}$ &$5.1\times 10^{-5}$ &$5.6\times 10^{-5}$\\
AL      &$3.4\times 10^{-4}$ &$5.3\times 10^{-4}$ &$7.6\times 10^{-4}$\\
CSM1    &$1.2\times 10^{-4}$ &$2.1\times 10^{-4}$ &$3.5\times 10^{-4}$\\
CSM2    &$1.6\times 10^{-3}$ &$4.4\times 10^{-3}$ &$1.2\times 10^{-2}$\\
\tableline
\end{tabular}
\end{center}
\end{table}

\newpage

\begin{figure}
\vspace{15cm}
\includegraphics{isernetal.ps1}
\noindent Figure 1: Luminosity functions corresponding to a burst with 
a  constant star formation rate that started at $T=13$ Gyr and lasted
$\Delta t = 0.1$ Gyr (upper solid line). The luminosity functions
obtained with the two methods, after binning in intervals of 1 
magnitude, are displayed below (they have been arbitrarely shifted 
for clarity). The differences between the usual method, solid 
line, and the direct method, dotted line, are very small.   
\end{figure}

\newpage

\begin{figure}
\vspace{17.5cm}
\includegraphics{isernetal.ps2a}
\includegraphics{isernetal.ps2b}
\noindent Figure 2: Luminosity functions of halo white dwarfs 
assuming bursts  of ages 10, 12, 14, 16 and 18 Gyr that lasted for 0.1 
Gyr as a function of the luminosity (panel a), and visual magnitude 
(panel b). In both cases, the luminosity function of disk white dwarfs 
(dashed line) for $t_{\rm disk}=9.3$ Gyr is also plotted for comparison 
purposes. The observational data were obtained from Liebert et al. 
(1988, 1989). 
\end{figure}

\newpage

\begin{figure}
\vspace{15cm}
\includegraphics{isernetal.ps3}
\noindent Figure 3: Luminosity functions for three bursts that 
started at $T=12$ Gyr and lasted 0.1, 1 and 3 Gyr. 
\end{figure}

\newpage

\begin{figure}
\vspace{17.5cm}
\includegraphics{isernetal.ps4}
\noindent Figure 4: Color--magnitude diagrams for the same bursts of 
figure 2.  
\end{figure}

\newpage

\begin{figure}
\vspace{15cm}
\includegraphics{isernetal.ps5}
\noindent Figure 5: Color--magnitude diagrams for a burst of age 12 
Gyr that lasted 3 Gyr.  
\end{figure}

\newpage

\begin{figure}
\vspace{15cm}
\includegraphics{isernetal.ps6}
\noindent Figure 6: Discovery function of halo white dwarfs in the 
$I$ band (upper panel) and in the $V$ band (lower panel) for bursts 
that started at 12, 14 and 16 Gyr and lasted 0.1 Gyr. In all cases, 
the limiting magnitude is $M_{V,I}=20$. 
\end{figure}

\newpage

\begin{figure}
\vspace{17.5cm}
\includegraphics{isernetal.ps7}
\noindent Figure 7: Comparison between the luminosity functions of
halo white dwarfs of ages 12 and 14 Gyr and different IMFs (see text
for details). 
\end{figure}

\end{document}